\documentclass[12pt]{article}
\usepackage{amsmath,color}
\usepackage[text={17cm,22.5cm}]{geometry}
\usepackage{ucs}
\usepackage[T1]{fontenc}
\usepackage[utf8]{inputenc}
\numberwithin{equation}{section}
\def\aff#1{\vspace{-12pt}{\normalsize #1}}
\def\lab#1{\label{eq:#1}}             \def\eq#1{(\ref{eq:#1})}
\def\bsubeq{\begin{subequations}}     \def\esubeq{\end{subequations}}
\def\nn{\nonumber}

\title{Black holes and naked singularities in four dimensional (A)dS Chamseddine gravity}

\author{B. Cvetkovi\'c  and D. Simi\'c\footnote{
        Email addresses: \texttt{cbranislav@ipb.ac.rs, dsimic@ipb.ac.rs}} \\
\aff{Institute of Physics, University of Belgrade,
                           Pregrevica 118, 11080 Belgrade, Serbia} }

\date{}
\begin{document}
\maketitle
	
\begin{abstract}
		We analyze solutions of Chamseddine's topological gravity in four space-time dimensions and discover various black hole solutions with(out) torsion as well as those that describe naked singularities. Because all of the solutions belong to the sector with vanishing scalar fields, they share peculiar trait that all conserved charges are vanishing.
	\end{abstract}

\section{Introduction}

Black holes (BH) are the most fascinating objects in Universe. They posses a number of remarkable properties that are, at the same time, guiding principles towards quantum theory of gravitation and the biggest unsolved problems in gravity.

A new chapter in our understanding of black holes and gravity as a whole started with the discovery of black hole thermodynamics. Besides energy, angular momentum and electric charge black holes have both entropy and temperature, both of which are very difficult  to understand from the  classical point of view. Hawking radiation \cite{a1} demonstrated that black holes are not really black and explained origin of it's temperature but opened a new problem, black hole information paradox. In general relativity (GR) entropy is proportional to the surface area of the event horizon, this is a well established result that is obtained both using Wald conserved charge approach \cite{a2} and Euclidean action \cite{a3}. Up to this day the microscopical understanding of BH entropy is elusive.

The fact that is not often emphasized is that properties of black holes depend crucially on the theory of gravity, consequently many impossible scenarios within GR are possible in other theory.
Black hole entropy in theories with higher curvature terms and torsion can have  different form, being  not simply proportional to the area of the event horizon \cite{sn,ax}, or with different proportionality factor than in GR \cite{mb1,mb2, ay}.

Topological theories of gravity are widely studied  in odd dimensions \cite{a4,a5,a6,a7} where many black hole solutions are found and their thermodynamic properties have been studied in great detail. On the other hand, Chamseddine's topological gravity \cite{a8} in even dimensions is not even remotely that well explored. To authors' knowledge, the only preliminary search for black hole solutions was done in Ref. \cite{a9} and Hamiltonian structure was studied in Ref. \cite{a10}. On the other hand, topological gravity in even dimensions was recently studied in the context of holography with number of interesting results \cite{a11}. With an idea of obtaining better understanding of four dimensional topological gravity we start preliminary analysis of its solutions, with the focus on the black holes.

The paper is organized as follows. In section 2 we review the basic properties of  Chamseddine's topological gravity in four dimensions.
In section 3 we analyze static solutions (with and without torsion) of dS Chamseddine's gravity: spherical, hyperbolic and planar.
Section 4 is dedicated to the AdS sector of the theory, in section 5 we analyze the conserved charges, while
section 6 is devoted to concluding remarks. Appendix contains a short discussion about solution under the horizon.

\section{Chamseddine topological gravity in four dimensions}

Topological gravity in even dimensions was introduced by Chamseddine \cite{a8}. A priori, it does not have vielbein as dynamical variable but only spin connection $\omega^{AB}$ and scalar field $\Phi^A$. The action in four dimensions is given by
\begin{equation}
	S=\epsilon_{ABCDE}F^{AB}F^{CD}\Phi^E,
\end{equation}
where $F^{AB}$ is curvature
\begin{equation}
	F^{AB}=d\omega^{AB}+\omega^A_{\ C}\omega^{CB}.
\end{equation}
Capital Latin indices $A,B,C...$ may correspond to Lorentz or (anti) de Sitter groups. When groups in question are (anti) de Sitter we can introduce vielbein as part of connection,
and we shall do this explicitly in the subsequent text. It is easy to find the field equations by varying the action with respect to basic dynamical variables.

Field equation for scalar field $\Phi^A$ is
\begin{equation}\lab{2.3}
	\epsilon_{ABCDE}F^{AB}F^{CD}=0.
\end{equation}
Equation of motion for spin connection is given by
\begin{equation}\lab{2.4}
	\epsilon_{ABCDE}F^{CD}D\Phi^E=0,
\end{equation}
where $D$ is exterior covariant derivative for the connection $\omega^{AB}$.

\subsection{de Sitter}

Let us firstly  analyze de Sitter group $SO(1,4)$, which represents the group of isometries of the metric $\rm diag(+,-,-,-,-)$.
We can separate indices $A$ into Lorentz $a$ and additional index $4$. With this splitting we can define vielbein $e^a$ and separate the scalar field into Lorentz vector scalar field $\phi^a$ and true scalar $\phi$
\begin{eqnarray}
\omega^{a4}=\frac{e^a}{\ell}\,,\qquad \Phi^a=\phi^a\,,\qquad \Phi^4=\phi\,.
\end{eqnarray}
Where $\ell$ is constant with dimension of length, in the rest of the paper we set $\ell=1$.
Curvature separates into Lorentz curvature and torsion
\bsubeq
\begin{eqnarray}
&&F^{ab}=R^{ab}+e^ae^b,\\
&&F^{a4}=T^a.
\end{eqnarray}
\esubeq
We also introduce the notation $\epsilon_{abcd4}=\epsilon_{abcd}$.
Field equations for scalar field $\Phi^A$ \eq{2.3} after introducing  Lorentz indices take the form
\bsubeq
\begin{eqnarray}
&&\epsilon_{abcd}(R^{ab}+e^ae^b)(R^{cd}+e^ce^d)=0\,,\\
&&\epsilon_{abcd}T^b(R^{cd}+e^ce^d)=0\,.
\end{eqnarray}
\esubeq
Field equation for spin connection \eq{2.4}  in Lorentz indices also splits into two equations
\bsubeq
\begin{eqnarray}
&&\epsilon_{abcd}(R^{bc}+e^be^c)(\nabla\phi^d-\phi e^d)=0\,,\\
&&\epsilon_{abcd}[(R^{cd}+e^ce^d)(d\phi-\phi_ie^i)-2T^c(\nabla\phi^d-\phi e^d)]=0\,.
\end{eqnarray}
\esubeq

\subsection{Anti de Sitter}

Anti de Sitter (AdS) group $SO(2,3)$ is isometry of the metric $diag(+,-,-,-,+)$.
As in the dS case we separate indices into Lorentz $a$ and additional denoted by $4$.
The curvature splits into Lorentz curvature and torsion
\bsubeq
\begin{eqnarray}
&&F^{ab}=R^{ab}-e^ae^b\,,\\
&&F^{a4}=T^a\,.
\end{eqnarray}
\esubeq
The equation of motion for scalar $\Phi^A$ \eq{2.3} again separates into two equations
\bsubeq
\begin{eqnarray}
&&\epsilon_{abcd}(R^{ab}-e^ae^b)(R^{cd}-e^ce^d)=0\,,\\
&&\epsilon_{abcd}T^b(R^{cd}-e^ce^d)=0\,.
\end{eqnarray}
\esubeq
The equation of motion for the spin connection \eq{2.4} in Lorentz indices is rewritten as two equations
\bsubeq
\begin{eqnarray}
&&\epsilon_{abcd}(R^{bc}-e^be^c)(\nabla\phi^d+\phi e^d)=0\,,\\
&&\epsilon_{abcd}[(R^{cd}-e^ce^d)(d\phi-\phi_ie^i)-2T^c(\nabla\phi^d+\phi e^d)]=0\,.
\end{eqnarray}
\esubeq

\section{dS Chamseddine gravity}

We shall analyze static solutions of Chamseddine's gravity with high symmetry, namely: spherical, hyperbolic and planar. All solutions that we obtain are in the sector where all scalars vanish
\begin{equation}
	\phi^a=\phi=0\\,
\end{equation}
we will not explicitly write that scalars are zero later in text when describing solutions.

Due to the very complicated nature of the equations of motion solutions that we have found are either with all scalars equal to zero or with at least one function remaining arbitrary, if some scalars are not zero. The solutions with non-zero scalars and without arbitrary functions  are left for further research.

For the simplification of the calculations we use xAct Mathematica package \cite{c1} that is available at \cite{c2}.

\subsection{Spherical}

Vielbein for static and spherically symmetric space-time is of the form
\begin{eqnarray}\lab{3.2}
e^0= B(r) dt\,,\qquad 	e^1= \frac{dr}{B(r)}\,,\qquad  e^2= r d\theta\,,\qquad 	e^3= r \sin \theta d\varphi\,.
\end{eqnarray}
The most general static and spherically symmetric ansatz for the connection is of the form
\begin{eqnarray}
&&\omega^{01}=F_0(r)dt+F_1(r)dr\,,\qquad \omega^{02}=F_2(r)d\theta\,,\qquad \omega^{03}=F_2(r)\sin\theta d\varphi\nn\\
&&\omega^{12}=F_3(r)d\theta\,,\qquad \omega^{13}=F_3(r)\sin\theta d\varphi\,,\qquad \omega^{23}=\cos\theta d\varphi\,.
\end{eqnarray}

Since we are not looking for the most general solution, we shall make further restrictions in the ansatz
\begin{equation}
	F_1(r)=F_3(r)=0.
\end{equation}
With this ansatz the solution for metric function is
\begin{equation}
	B(r)=r \sqrt{\frac{r^2}{r_h^2}-1},
\end{equation}
and non-zero spin connection functions are
\begin{equation}
F_0(r)=-r\,\quad F_2(r)=\pm\frac{\sqrt{r^4-r^2+k}}{r},
\end{equation}
where $r_h$ and $k$ are constants, with the condition that $k\neq 0$. If $k=0$ then we are left with only one independent equation that determines $\phi^1(r)$ in terms of $B(r)$, consequently for $k=0$ solution is not uniquely determined.

The non-zero components of  torsion are given by:
\begin{eqnarray}\lab{3.31}
&&T^0=-\frac{2B(r)}re^0e^1\,,\qquad T^1=0\,,\nn\\
&&T^2=B(r)e^1e^2+F_2(r)e^0e^2\,,\qquad  T^3=B(r)e^1e^3+F_2(r)e^0e^3\,,
\end{eqnarray}
while Cartan curvature reads:
\begin{eqnarray}\lab{3.32}
&&R^{01}=e^0e^1\,,\qquad  R^{02}=\frac{B(r)F_2(r)'}re^1e^2\,,\nn\\
&&R^{12}=\frac{F_2(r)}{B(r)}e^0e^2\,,\qquad R^{03}=\frac{B(r)F_2(r)'}re^1e^3\,,\nn\\
&&R^{12}=\frac{F_2(r)}{B(r)}e^0e^3\,,\qquad R^{23}=-\frac{F_2(r)^2+1}{r^2}e^2e^3\,,
\end{eqnarray}
where $F_2(r)':=dF_2(r)/dr$. Then Cartan curvature scalar is:
\begin{equation}
R=2\left(2-r^2-\frac{k+1}{r^2}\right)\,,
\end{equation}
and it is divergent for $r=0$.
The quadratic curvature invariant reads:
\begin{equation}\lab{3.33}
R^{ijkl}R_{ijkl}=2r^2\left(1+\frac{(1+F_2(r)^2)^2}{r^4}-\frac{2F_2(r)^2}{B(r)^2}-\frac{2B(r)^2(F_2(r)')^2}{r^2}\right)\,,
\end{equation}
and it is also divergent for $r=0$.
Important point is that quadratic curvature invariant is a priori divergent at the horizon $r=r_h$,  for arbitrary $r_h$ and $k$. By demanding that it is finite at the horizon we determine value of the constant $k$ as function of location of the horizon $r_h$
\begin{equation}
	k=r_h^2-r_h^4.
\end{equation} 
Finally, Ricci scalar of the Riemannian curvauture  is given by
\begin{equation}
\tilde R=-12-\frac{2}{r}+\frac{30r^2}{r_h^2},
\end{equation}
and it is divergent at $r=0$, meaning that this solution describes spherically symmetric black hole. It is worth  noting, that in order to obtain Ricci scalar, we continued metric inside  the horizon, as it is normally done when studding BH solutions.

The spin connection is singular inside of the horizon, because  the square root becomes complex for $r<r_h$. We note that the same holds for metric function $B(r)$, which simply means that obtained solution is valid only outside of the horizon, i.e. for $r\geq r_h$. Consequently, we only need to demand that spin connection is real outside of the horizon. Solution inside of the horizon is discussed in appendices.

\subsection{Hyperbolic}

Vielbein is given by
\begin{eqnarray}\lab{3.10}
e^0= B(r) dt\,, \qquad	e^1= \frac{dr}{B(r)}\,,\qquad  e^2= r d\theta\,,\qquad 	e^3= r \sinh\theta d\varphi\,.
\end{eqnarray}
Hyperbolic connection takes form
\begin{eqnarray}
&&\omega^{01}=F_0(r)dt+F_1(r)dr\,,\qquad \omega^{02}=F_2(r)d\theta\,,\qquad \omega^{03}=F_2(r)\sinh\theta d\varphi\,,\nn\\
&&\omega^{12}=F_3(r)d\theta\,,\qquad \omega^{13}=F_3(r)\sinh\theta d\varphi\,,\qquad \omega^{23}=\cosh\theta d\varphi\,.
\end{eqnarray}
The metric function is the same as in spherically symmetric case
\begin{equation}
	B(r)=r \sqrt{\frac{r^2}{r_h^2}-1}\,.
\end{equation}
Non-zero spin connection functions are
\begin{equation}
F_0(r)=-r\,,\quad F_2(r)=\pm\frac{\sqrt{r^4+r^2+k}}{r},
\end{equation}
where $r_h$ and $k$ are constants and $k\neq 0$ for the same reason as in spherical case.

Torsion and Cartan curvature are given by the same expressions as in spherical case \eq{3.31} and \eq{3.32}, only vielbein and spin connection are changed by their hyperbolic values. Quadratic curvature invariant is of the same form as in spherical \eq{3.33} and from requirement that it is regular at the horizon we obtain value of the constant $k$
\begin{equation}
	k=-r_h^2-r_h^4.
\end{equation}
Ricci scalar is given by
\begin{equation}
	\tilde{R}=-12+\frac{2}{r}+\frac{30r^2}{r_h^2},
\end{equation}
and is divergent at the origin, proving that this solution describes black hole with hyperbolic symmetry.
\subsection{Planar}
Vielbein with planar symmetry is given by
\begin{eqnarray}\lab{3.17}
e^0= B(r) dt\,,\quad 	e^1=\frac{dr}{B(r)}\,,\quad  e^2= r dx\,,\quad  	e^3= r dy\,.
\end{eqnarray}
Non-zero components of spin connection are assumed to be of the same form as in previous cases
\begin{eqnarray}
\omega^{01}=F_0(r)dt\,,\qquad \omega^{02}=F_2(r)dx\,,\qquad \omega^{03}=F_2(r) dy\,.
\end{eqnarray}
Metric function is, again, the same
\begin{equation}
	B(r)=r \sqrt{\frac{r^2}{r_h^2}-1},
\end{equation}
while spin connection functions are
\begin{equation}
	F_0(r)=-r,\ F_2(r)=\pm\frac{\sqrt{r^4+k}}{r},
\end{equation}
where $r_h$ and $k$ are constants. The crucial point, as before, is that $k$ is non-vanishing.
Torsion and Cartan curvature are, again, are of the same form as in spherical case \eq{3.31} and \eq{3.32}, with vielbein and spin connection taking values for planar case. Quadratic curvature invariant is the same as in spherical case \eq{3.33} and from requirement that it is regular at the horizon we obtain value of the constant $k$
\begin{equation}
	k=-r_h^4.
\end{equation}
The Cartan curvature scalar
\begin{equation}
	R=-\frac{k}{r^4},
\end{equation}
and is divergent for $r=0$.

The Ricci scalar is regular
\begin{equation}
	\tilde{R}=-12+\frac{30r^2}{r_h^2}.
\end{equation}
Because everything is regular at the surface $r=r_h$ it represents coordinate singularity combined with divergence of the Cartan curvature scalar at $r=0$ shows that this solution represents black hole with planar horizon.
\subsection{Solutions without torsion}

The form of  the spherical \eq{3.2}, hyperbolic \eq{3.10} and planar \eq{3.17} vielbein is the same as before.
Torsion zero condition, $T^i=0$, determines the value of the spin connection in terms of metric function, and the only equation of motion that is not identically satisfied leads to the solution for the metric function
\begin{equation}
	B(r)=\sqrt{1-r^2\pm \sqrt{a+br}}.
\end{equation}
The function $B(r)$ has only one real zero and, since for the large values of $r$ the term $-r^2$ dominates, this zero represents  cosmological horizon. The spin connection is given by
\bsubeq
\begin{eqnarray}
&&F_1=F_2=0\,,\\
&&F_0=-B(r) B'(r)=r\mp\frac{b}{4\sqrt{a+br}}\,,\\
&&F_3(r)=B(r)=\sqrt{1-r^2\pm \sqrt{a+br}}\,.
\end{eqnarray}
\esubeq
The reality condition for the metric at $r=0$ implies
\begin{equation}
	a\geq0\\,
\end{equation}
while for the choice of minus sign in $B(r)$ we, also, must have that $a$ is not greater than one. In the spherically symmetric case Ricci scalar $R$ diverges at $r=0$
\begin{equation}
	R=-12 \pm \frac{8a^2+24abr+15b^2r^2}{4r^2(a + b r)^{3/2}},
\end{equation}
unless both $a$ and $b$ are zero. Consequently, this solution represent a naked singularity, except when for $a=b=0$ it reduces to dS spacetime.

In the hyperbolic case Ricci scalar is divergent even when $a=b=0$
\begin{equation}
	R=-12 \mp\frac{b^2}{4(a + b r)^{3/2}}\pm\frac{2(a+2br+2\sqrt{a+br})}{r^2\sqrt{a+br}}.
\end{equation}
In the planar case the situation is the same as in the hyperbolic case
\begin{equation}
	R=-12 \mp\frac{b^2}{4(a + b r)^{3/2}}+\frac{2(\pm a\pm2br+2\sqrt{a+br})}{r^2\sqrt{a+br}}\,,
\end{equation}
the solution has naked singularity at $r=0$ even when $a=b=0$.

\section{AdS Chamseddine gravity}

\subsection{Black hole solutions without torsion}
We use the same ansatz for vielbein as in dS case, namely \eq{3.2} for spherical, \eq{3.10} for hyperbolic and \eq{3.17} for planar.

The only equation of motion that is not identically satisfied gives the solution for metric function
\begin{equation}
	B(r)=\sqrt{1+r^2\pm \sqrt{a+br}}.
\end{equation}
Torsion zero condition gives the value of spin connection
\bsubeq
\begin{eqnarray}
&&F_1(r)=F_2(r)=0\,,\\
&&F_0(r)=-B(r) B'(r)=-r\mp\frac{b}{4\sqrt{a+br}}\,,\\
&&F_3(r)=B(r)\sqrt{1+r^2\pm \sqrt{a+br}}\,.
\end{eqnarray}
\esubeq
For the choice of minus sign in the metric function $B$ we can get black hole solutions with one or two horizons as well as naked singularities. Presence of two parameters allows for tuning so that horizon appears. For the plus sign for all values of parameters $a$ and $b$ solutions represent naked singularities.

In spherically symmetric case Ricci scalar for the Riemannian curvature $\tilde R$ diverges at $r=0$
\begin{equation}
\tilde R=12 \pm \frac{8a^2+24abr+15b^2r^2}{4r^2(a + b r)^{3/2}},
\end{equation}
unless both $a$ and $b$ are zero, which describes AdS spacetime.

In the hyperbolic case Ricci scalar is divergent even when $a=b=0$
\begin{equation}
\tilde R=12 \mp\frac{b^2}{4(a + b r)^{3/2}}\pm\frac{2(a+2br+2\sqrt{a+br})}{r^2\sqrt{a+br}}\,.
\end{equation}
In planar case we encounter the same situation as in the hyperbolic case
\begin{equation}
	\tilde R=12 \mp\frac{b^2}{4(a + b r)^{3/2}}+\frac{2(\pm a\pm2br+\sqrt{a+br})}{r^2\sqrt{a+br}}\,,
\end{equation}
i.e. the solution has a singularity at $r=0$, even when $a=b=0$.

Solutions with the minus sign in front of the square root can have one or two horizons, depending on the values of parameters $a$ and $b$, as we checked for some choices.

\subsection{Solutions with torsion}

We use the same ansatz for vielbein and spin connection as in dS case.

The solution for the metric function reads
\begin{equation}
	B(r)=r \sqrt{1\pm\frac{r^2}{r_h^2}},
\end{equation}
while non-zero spin connection functions are
\begin{equation}
F_0(r)=r\,,\qquad F_2(r)=\pm\frac{\sqrt{-r^4+sr^2+k}}{r}\,,
\end{equation}
where $r_h$ and $k$ are constants and $s$ is equal to $0$ for planar, +1 for hyperbolic and $-1$ for spherical case. The crucial point, as before, is that we must have $k\neq 0$ to obtain a unique solution. Signs in  the metric function $B$ and spin connection function $F_2$ are independent, i.e for any choice of signs there is a solution.

In spherical and hyperbolic case Ricci scalar $R$ is divergent at $r=0$
\begin{eqnarray}
R_{Spherical}=12-\frac{2}{r^2}\pm 30\frac{r^2}{r_h^2},\\ R_{Hyperbolic}=12+\frac{2}{r^2}\pm 30\frac{r^2}{r_h^2},
\end{eqnarray}
meaning that this solutions describe naked singularity.
In the planar case Ricci scalar is regular everywhere
\begin{equation}
		R=12\pm 30\frac{r^2}{r_h^2}\,.
\end{equation}

\section{Charges}
\setcounter{equation}{0}
The simplest and the shortest way to determine charges in the case when all scalars are equal to zero is by  using the covariant phase space formalism \cite{e1}.

The variation of the action reads
\begin{equation}
	\delta S=\int E.O.M+d\left(\epsilon_{ABCDE}\delta\omega^{AB}F^{CD}\Phi^E\right)\\,
\end{equation}
from which we determine the presymplectic potential current
\begin{equation}
\theta=\epsilon_{ABCDE}\delta\omega^{AB}F^{CD}\Phi^E\\.
\end{equation}
In the sector of the theory where all scalars are equal to zero presymplectic potential current vanishes
\begin{equation}
	\theta=0\\.
\end{equation}
As a direct consequence, presymplectic form current is also zero in this sector
\begin{equation}
\omega=\delta\theta=0\\.
\end{equation}
This implies that all the charges are zero when all scalars vanish, as well as that the number of degrees of freedom is zero in this sector. For explicit calculations of charges from symplectic form see for example \cite{e2}.

\section{Concluding remarks}
We investigated solutions of Chamseddine's topological gravity in four dimensions. Due to very complicated nature of the equations of motion we found only solutions for which all scalar fields are vanishing. This corresponds to trivial sector of the theory because all charges and degrees of freedom vanish. Although being trivial from the point of view of the charges, this sector has a rich solution space.

We focused on highly symmetric solutions in order to simplify equations as much as possible and obtained static solutions with spherical, hyperbolic and planar symmetry. The solutions with torsion in de Sitter case are black holes while solutions without torsion describe naked singularities. On the contrary in anti de Sitter case solutions without torsion represent black holes.

Black holes in dS Chamseddine gravity had a priory arbitrary non-zero constant $k$ in spin connection. By demanding regularity of the solution at $r=r_h$, from quadratic Cartan curvature invariant we determined the value of $k$ in term of $r_h$. Only the solutions with that special value of $k$ describe a black hole, otherwise it is singular at $r=r_h$.
Also, BH in dS Chamseddine gravity have a peculiar property that one cannot take radius of the horizon into zero and obtain regular solutions that are not a black hole.

More general solutions with non-zero scalar fields, that correspond to the more interesting sector of the theory, are left for further studies.

\section*{Acknowledgements}

This work was partially supported by the Ministry of  Science, Technological development and Inovations of the Republic of Serbia.\\
We thank Danilo Rakonjac for useful comments and suggestions.

\appendix
\section{Solution inside of the horizon}

\subsection{dS black holes}
We search for a solution inside of the horizon by first extending metric $g_{\mu\nu}=e^a_\mu e^b_\nu\eta_{ab}$ to have the same functional form  inside of the horizon as it has outside.
Because metric components $g_{tt}$ and $g_{rr}$ change signs inside of the horizon, effectively $r$ becomes time coordinate and $t$ space coordinate.

Vielbein inside of the horizon $r<r_h$ is determined from the metric inside of the horizon in the standard way. Vielbeins $e^2$ and $e^3$ stay the same as outside of the horizon, while $e^0$ and $e^1$ are given by
\begin{eqnarray}
	e^0=\frac{dr}{r\sqrt{1-\frac{r^2}{r_h^2}}}, & e^1=r\sqrt{1-\frac{r^2}{r_h^2}}dt.
\end{eqnarray}
Form of the spin connection remains the same as the ansatz for connection, while the value of functions are obtained by demanding the same form of torsion as outside of the horizon. The solution that fulfills this requirements is 
\begin{eqnarray}
	F_0(r)=r, & F_3(r)=\mp\frac{\sqrt{-r^4-sr^2-k}}{r},
\end{eqnarray}
where $s$ is $\pm1$ or zero depending on the topology of the horizon, as used previously in the text.

One could have guessed the result by noticing that inside of the horizon we changed places of $e^1$ and $e^0$, effectively making a replacement of indices $0\leftrightarrow1$. With the addition that functions under square root should change sign to become positive the same way as in vielbein. 
\subsection{AdS Black holes}
Vielbein inside of the horizon $r<r_h$ is
\begin{eqnarray}
	e^0=\frac{dr}{\sqrt{-1-r^2+\sqrt{a+br}}}, & e^1=\sqrt{-1-r^2+\sqrt{a+br}}dt,
\end{eqnarray}
while $e^2$ and $e^3$ remain the same as outside of the horizon.
Condition that torsion is zero determines spin connection functions
\begin{eqnarray}
	F_0(r)=r-\frac{b}{4\sqrt{a+br}} & F_2(r)=-\sqrt{-1-r^2+\sqrt{a+br}}
\end{eqnarray}
Notice that we assume that values of $a$ and $b$ are such that there is only one horizon. As we already said, for some choices there is no horizon at all or there can even be two. 

\end{document}